\documentclass[12pt]{iopart}

\usepackage{iopams}
\usepackage{graphicx}
\begin{document}

\title[Natural time and 1/f ``noise'']{Natural time and 1/f ``noise''}

\author{P A Varotsos}
\address{Solid State Section  and Solid Earth Physics Institute, Physics Department, University of Athens, Panepistimiopolis, Zografos 157 84,
Athens, Greece}
\ead{pvaro@otenet.gr}
\author{N V Sarlis}
\address{Solid State Section  and Solid Earth Physics Institute, Physics Department, University of Athens, Panepistimiopolis, Zografos 157 84,
Athens, Greece}
\author{E S Skordas}
\address{Solid State Section  and Solid Earth Physics Institute, Physics Department, University of Athens, Panepistimiopolis, Zografos 157 84,
Athens, Greece}

\begin{abstract}
Seismic electric signals have been found to obey the ubiquitous
$1/f^a$ behavior [{\em Phys. Rev. E} {\bf 66}, 011902(2002)]. The
newly introduced concept of natural time enables the study of the
dynamic evolution of a complex system and identifies when the
system enters the critical stage. On the basis of this concept, a
simple model is proposed here which exhibits the $1/f^a$ behavior
with $a$ close to unity. Furthermore, we present recent data of
electric signals, which when analyzed in the natural time domain
are found to exhibit {\em critical} dynamics and hence can be
classified as seismic electric signals.
\end{abstract}

\pacs{05.40.-a, 05.45.Tp, 89.75.-k, 89.75.Da}
\maketitle

\section{Introduction}
Among the different features that characterize complex physical
systems, the most ubiquitous is the presence of $1/f^a$ noise in
fluctuating physical variables\cite{MAN99}. This means that the
Fourier power spectrum $S(f)$ of fluctuations scales with
frequency $f$ as $S(f) \sim 1/f^a$. The power-law behavior often
persists over several orders of magnitude with cutoffs present at
both high and low frequencies. Typical values of the exponent $a$
approximately range between 0.8 and 4 (e.g., see Ref.\cite{ANT02}
and references therein), but in a loose terminology   all these
systems are said to exhibit $1/f$ ``noise''. Such a ``noise'' is
found in a large variety of systems, e.g., condensed matter
systems(e.g. \cite{WEI88}), freeway
traffic\cite{MUS76,NAG95,ZHA95}, granular flow\cite{NAK97}, DNA
sequence\cite{gol02}, heartbeat\cite{PEN93}, ionic current
fluctuations in membrane channels\cite{MER99}, river
discharge\cite{MAN69b}, the number of stocks traded
daily\cite{LIL00}, chaotic quantum
systems\cite{GOM05,REL02,SAN05,SAN06}, the light of
quasars\cite{PRE78}, human cognition\cite{GIL95} and
coordination\cite{YOS00}, burst errors in communication
systems\cite{BER63}, electrical measurements\cite{KOG96}, the
electric noise in carbon nanotubes\cite{COL00} and in nanoparticle
films\cite{KIS97}, the occurrence of earthquakes\cite{SOR00} etc.
In some of these systems, the exponent $a$ was reported to be very
close to 1, but good quality data  supporting  such a value exist
in a few of them\cite{WEI88}, e.g., the voltage fluctuations when
current flows through a resistor\cite{YAK00}.

The $1/f^a$ behavior has been well understood on the basis of
dynamic scaling observed at {\em equilibrium} critical points
where the power-law correlations in time stem from the
infinite-range correlations in space (see Ref.\cite{ANT02} and
references therein). Most of the observations mentioned above,
however, refer to {\em nonequilibrium} phenomena for which
-despite some challenging theoretical
attempts\cite{BAK87,BAK96,ANT01,DAV02}- possible {\em generic}
mechanisms leading to scale invariant fluctuations have not yet
been identified. In other words, despite its ubiquity, there is no
yet universal explanation about the phenomenon of the $1/f^a$
behavior. Opinions have been expressed (e.g., see
Ref.\cite{GOM05}) that it does not arise as a consequence of
particular physical interactions, but it is a generic
manifestation of complex systems.

It has been recently
shown\cite{NAT01,NAT02,NAT02A,NAT03,NAT03B,NAT04,NAT05} that novel
dynamic features hidden behind the time series of complex systems
can emerge if we analyze them in terms of a newly introduced time
domain, termed natural time $\chi$. It seems that this analysis
enables the study of the dynamic evolution of a complex system and
identifies when the system enters a critical stage.  Natural time
domain is optimal\cite{ABE05} for enhancing the signal's
localization in the time frequency space, which  conforms to the
desire to reduce uncertainty and extract signal information as
much as possible. The scope of the present paper is twofold.
First,  a simple model is proposed (Section 2) which, in the frame
of natural time analysis, leads to $1/f^a$ behavior with an
exponent close to unity. Second, we present the most recent
experimental data on Seismic Electric Signals (SES) which are
transient low frequency ($\leq 1Hz$) signals  observed before
earthquakes \cite{proto,var86b,var88x,var99,grl}, since they are
emitted when the stress in the focal region reaches a critical
value before the failure\cite{varbook}. The analysis of the
original time series of the SES activities have been shown to obey
a $1/f$-behavior\cite{NAT02}. Here in Section 3, we show that the
SES activities observed in Greece during the last months exhibit
the features suggested, on the basis of natural time, to describe
critical dynamics.
 Such features have been found\cite{VAR06PRB} for laboratory  data\cite{AEG03,AEG04A} of the
avalanches in a three dimensional pile of rice (which is similar
to the prototype example of sandpiles used in the proposal of the
Self-Organized Criticality, SOC\cite{BAK87,BAK96}) getting
progressively closer to the critical state.

In a time series comprising $N$ events, the {\em natural time}
$\chi_k = k/N$ serves as an index\cite{NAT01,NAT02,NAT02A} for the
occurrence of the $k$-th event. The evolution of the pair
($\chi_k, Q_k$) is
studied\cite{NAT01,NAT02,NAT02A,NAT03,NAT03B,NAT04,NAT05,NAT05B,newbook,VAR05C,NAT06A},
where $Q_k$ denotes a quantity proportional to the energy released
in the $k$-th event. For example, for dichotomous signals, which
is frequently the case of SES activities, $Q_k$ stands for the
duration of the $k$-th pulse.
 The normalized power spectrum $\Pi(\omega )\equiv | \Phi (\omega ) |^2 $ was
introduced\cite{NAT01,NAT02}, where
\begin{equation}
\label{eq3} \Phi (\omega)=\sum_{k=1}^{N} p_k \exp \left( i \omega
\frac{k}{N} \right)
\end{equation}
and $p_k=Q_{k}/\sum_{n=1}^{N}Q_{n}$, $\omega =2 \pi \phi$; $\phi$
stands for the {\it natural frequency}. The continuous function
$\Phi (\omega )$ should {\em not} be confused with the usual
discrete Fourier transform, which considers only its values at
$\phi=0,1,2,\ldots$. In natural time analysis, the properties of
$\Pi(\omega)$ or $\Pi(\phi)$ are studied (\cite{NAT01,NAT02}) for
natural frequencies $\phi$
 less than 0.5, since in
this range of $\phi$, $\Pi(\omega)$  or $\Pi(\phi)$ reduces
(\cite{NAT01,NAT02,NAT02A,newbook})
 to a {\em characteristic function} for the
probability distribution $p_k$  in the context of probability
theory.
 When the system enters the
{\em critical} stage, the following relation
holds\cite{NAT01,NAT02,VAR05C}:
\begin{equation}
\Pi ( \omega ) = \frac{18}{5 \omega^2} -\frac{6 \cos \omega}{5
\omega^2} -\frac{12 \sin \omega}{5 \omega^3}. \label{fasma}
\end{equation}
For $\omega \rightarrow 0$, Eq.(\ref{fasma}) leads
to\cite{NAT01,NAT02,newbook}
\[ \Pi (\omega )\approx 1-0.07
\omega^2\] which reflects\cite{VAR05C} that the variance of $\chi$
is given by
\[ \kappa_1=\langle \chi^2 \rangle -\langle \chi \rangle
^2=0.07,\] where $\langle f( \chi) \rangle = \sum_{k=1}^N p_k
f(\chi_k )$.
  The entropy $S$ in the natural time-domain is defined
as\cite{NAT01,NAT03B} \[ S \equiv  \langle \chi \ln \chi \rangle -
\langle \chi \rangle \ln \langle \chi \rangle,\] which  depends on
the sequential order of events\cite{NAT04,NAT05} and for {\em
infinitely} ranged temporal correlations  its value is
smaller\cite{NAT03B,newbook} than the value $S_u (=\ln 2
/2-1/4\approx 0.0966$) of a ``uniform'' distribution (as defined
in Refs. \cite{NAT01,NAT03,NAT03B,NAT04,NAT05}, e.g.   when all
$p_k$ are equal or $Q_k$ are positive indepedent and identically
distributed random variable of finite variance), i.e.,$S < S_u$.
The same holds for the value of the entropy obtained\cite{NAT05B}
upon considering the time reversal ${\mathcal T}$, i.e.,
${\mathcal T} p_k=p_{N-k+1}$, which is labelled by $S_-$. In
summary, the SES activities, when analyzed in natural time exhibit
{\em infinitely} ranged temporal correlations and  obey the
conditions:
\begin{equation}\label{eq1}
    \kappa_1 = 0.07
\end{equation}
and
\begin{equation}\label{eq2}
    S, S_- < S_u.
\end{equation}
The validity of these conditions for the most recent electric
field data will be investigated in Section 3.

\begin{figure*}
\includegraphics{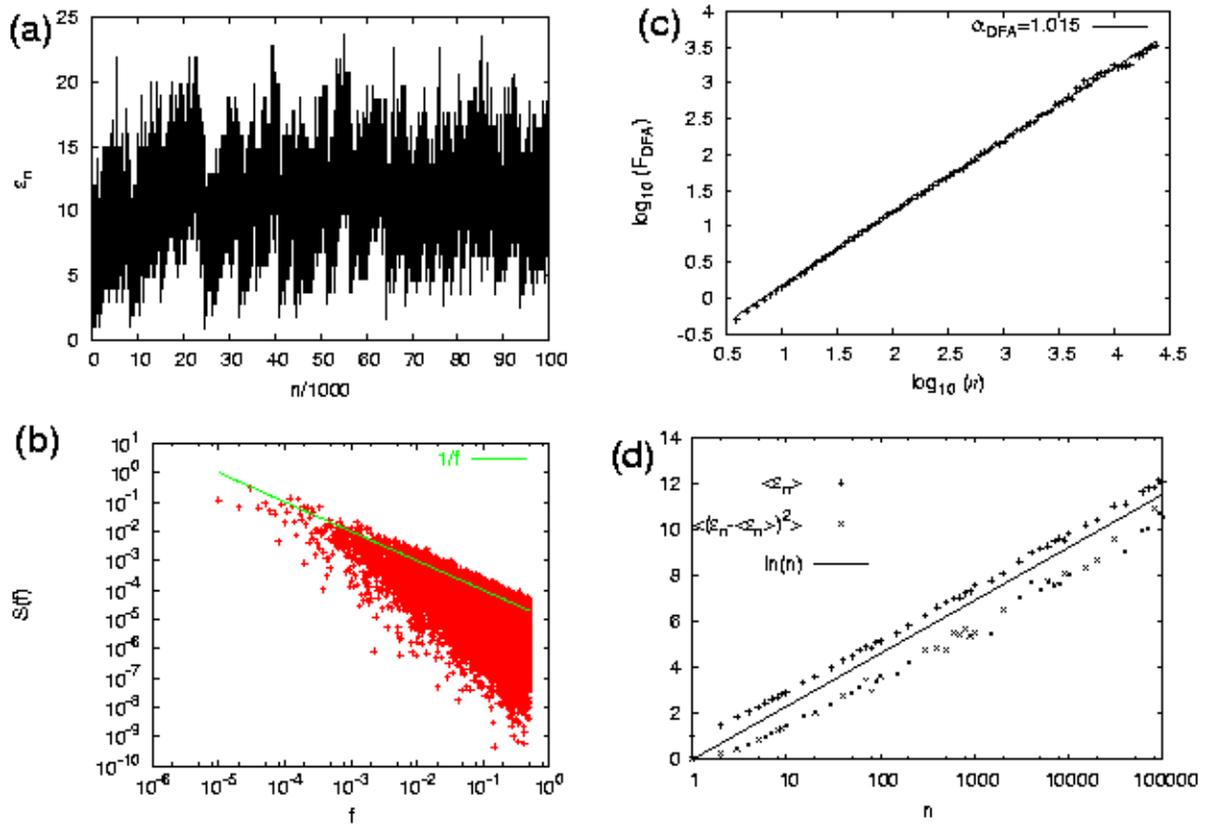}
\caption{(color) (a):Example of the evolution of
$\epsilon_n$ (see the text) versus the number of renewals $n$,
i.e., in natural time. (b): The Fourier  power spectrum of (a);
the (green) solid line corresponds to $1/f$ and was drawn as a
guide to the eye. (c): The DFA of (a) that exhibits an exponent
$\alpha_{DFA}$ very close to unity, as expected from (b).
(d):Properties of the distribution of $\epsilon_n$ that explain
the small deviation of $\alpha_{DFA}$ from unity. The average
value $\langle \epsilon_n \rangle$ (plus) and the variance
$\langle \left( \epsilon -\langle
 \epsilon_n \rangle \right)^2 \rangle$ (crosses) as a function of n. The straight solid line depicts $\ln (n)$ and was drawn for the sake of reader's convenience.} \label{fmodel}
\end{figure*}

\begin{figure*}
 \includegraphics{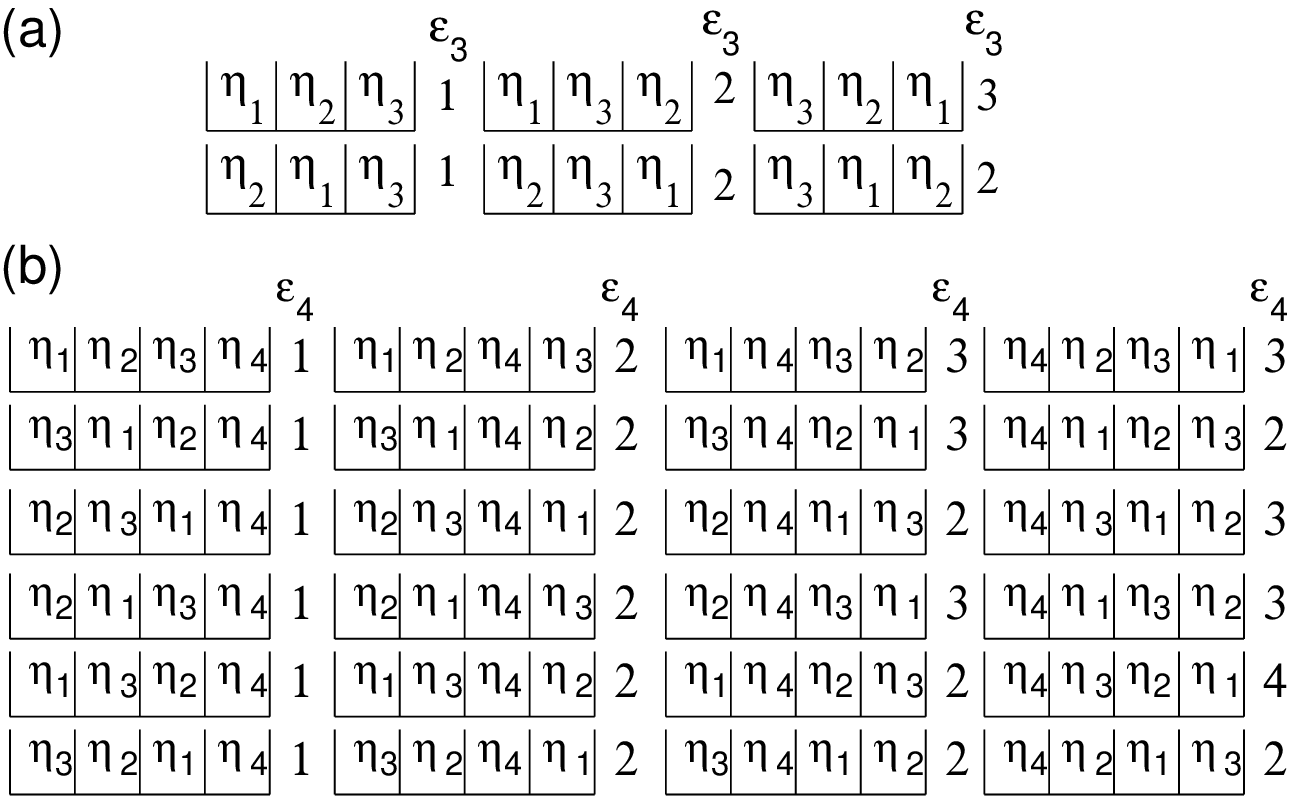}
 \caption{(a)The six(=3!) equally probable outcomes after the selection of 3 random numbers by the same PDF. Actually, the sample space is (in one to one correspondence to) the permutations of 3 objects. (b) The 24(=4!) equally probable outcomes after the selection of 4 random numbers by the same PDF. Again, the sample space is (in one to one correspondence to) the permutations of 4 objects. For the readers convenience, in each outcome, the corresponding $\epsilon_n$-value  ($n=3$ or 4)is written. An inspection of (b), shows that $p(\epsilon_4=1)=1/4, p(\epsilon_4=2)=11/24,  p(\epsilon_4=3)=1/4$ and $p(\epsilon_4=4)=1/24$. }
 \label{x4}
\end{figure*}

\section{The model proposed}

Here, we present a  simple competitive evolution model which
results, when analyzed in natural time, to $1/f^a$ ``noise'' with
$a$ very close to unity. Let us consider the cardinality
$\epsilon_n$ of the family of sets ${\rm S_n}$ of successive
extrema obtained from a given probability distribution function
(PDF); ${\rm S_0}$ equals to the empty set. Each ${\rm S_n}$ is
obtained by following the procedure described below for n times.
Select a random number $\eta_n$ from a given PDF and compare it
with all the numbers of ${\rm S_{n-1}}$.  In order to construct
the set ${\rm S_{n}}$,  we disregard from the set ${\rm S_{n-1}}$
all its members that are smaller than $\eta_n$ and furthermore
include $\eta_n$. Thus, ${\rm S_{n}}$  is a finite set of real
numbers whose members are always larger or equal to  $\eta_n$.
Moreover $\min[ {\rm S_n}] \geq \min[{\rm S_{n-1}}]$ and $\max
[{\rm S_n}] \geq \max[{\rm S_{n-1}}]$. The cardinality $\epsilon_n
\equiv \left| {\rm  S_n }\right|$ of these sets, which may be
considered as equivalent to the dimensionality of the thresholds
distribution in the coherent noise model (e.g. see
Ref.\cite{ABE04} and references therein), if considered as
time-series with respect to the natural number n (see
Fig.\ref{fmodel}(a), which was drawn by means of the exponential
PDF) exhibits $1/f^a$ noise with $a$ very close to unity, see
Fig.\ref{fmodel}(b). This very simple model whose evolution is
depicted in Fig.\ref{fmodel}(a), leads to a detrended fluctuation
analysis\cite{PEN93} (DFA) exponent $\alpha_{DFA} \approx 1.02$,
see Fig.\ref{fmodel}(c), being compatible with the $1/f$ power
spectrum depicted in Fig.\ref{fmodel}(b). The mathematical model
described above corresponds to an asymptotically non-stationary
process, since $\langle \epsilon_n \rangle \propto \ln n$ with a
variance $\langle \left( \epsilon -\langle \epsilon_n \rangle
\right)^2 \rangle \propto \ln n$ (see Fig.\ref{fmodel}(d)), and
this conforms to the fact that the DFA exponent is slightly larger
than unity.

We now discuss an analytical procedure which clarifies some
properties of the model. In order to  find  analytically the
distribution of the probabilities $p(\epsilon_n)$, one has simply
to consider the possible outcomes when drawing n random numbers
$\eta_n$. Since the selection is made by a means of a PDF, all
these numbers are different from each other, thus -when sorted
they- are equivalent to  n points (sites) lying on the real axis.
The value of $\epsilon_n$ varies as $\left\{ \eta_n \right\}$
permutate along these  n sites {\em independently} from the PDF
used in the calculation. Thus, a detailed study of the permutation
group of n objects can lead to an exact solution of the model. It
is well known, however, that the number of the elements of this
group is n! and this explains why we preferred to use the
numerical calculation shown in Fig.\ref{fmodel}.  Some exact
results obtained by this method are the following: $\langle
\epsilon_1 \rangle=1$; $\langle \epsilon_2 \rangle=1+1/2$, since
$p(\epsilon_2=1)=p(\epsilon_2=2)=1/2$; $\langle \epsilon_3
\rangle=1+1/2+1/3$, since $p(\epsilon_3=1)=1/3,
p(\epsilon_3=2)=1/2$ and $p(\epsilon_3=3)=1/6$; $\langle
\epsilon_4 \rangle=1+1/2+1/3+1/4$ (see Fig.\ref{x4}). Figure
\ref{x4} analyzes the results for $n=3$ (Fig.\ref{x4}(a)) and
$n=4$ (Fig.\ref{x4}(b)).  One can see that the probability
$p(\epsilon_n=m)$ equals to the sum of the $n$ possible outcomes
as $\eta_n$ moves from the left to right in the $n$ columns of
Fig.\ref{x4}. In each column, the probability to have at the end
$\epsilon_n=m$ is just equal to the probability to keep  $m-1$
numbers from the numbers already drawn that are larger than
$\eta_n$. This results in
\begin{equation}
p(\epsilon_n=m)=\frac{1}{n} \sum_{k=m-1}^{n-1} p(\epsilon_k=m-1).
\label{dom}
\end{equation}
Using Eq.(\ref{dom}), one can prove that $\langle \epsilon_n
\rangle =\langle \epsilon_{n-1} \rangle+1/n$, which reflects that
$\langle \epsilon_n \rangle =\sum_{k=1}^n1/k$.

\section{The recent electric field data}

\begin{figure*}
\includegraphics{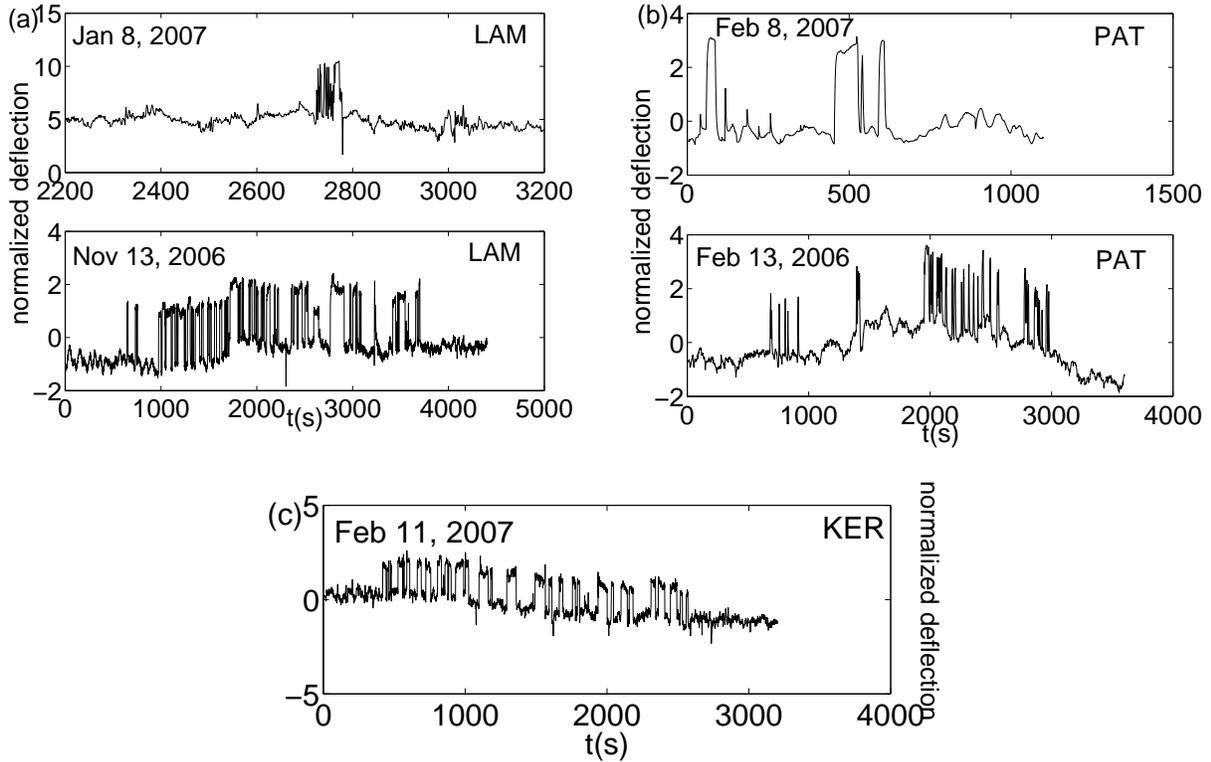}
\caption{Electric signals recorded at LAM (a), PAT (b) and KER
(c). (sampling rate $f_{exp}$=1 sample/sec). The actual electric
field amplitude $E$ (in $mV/km$) in (c) is appreciably smaller
than that in (a) and (b).} \label{f3}
\end{figure*}

\begin{figure*}
\includegraphics{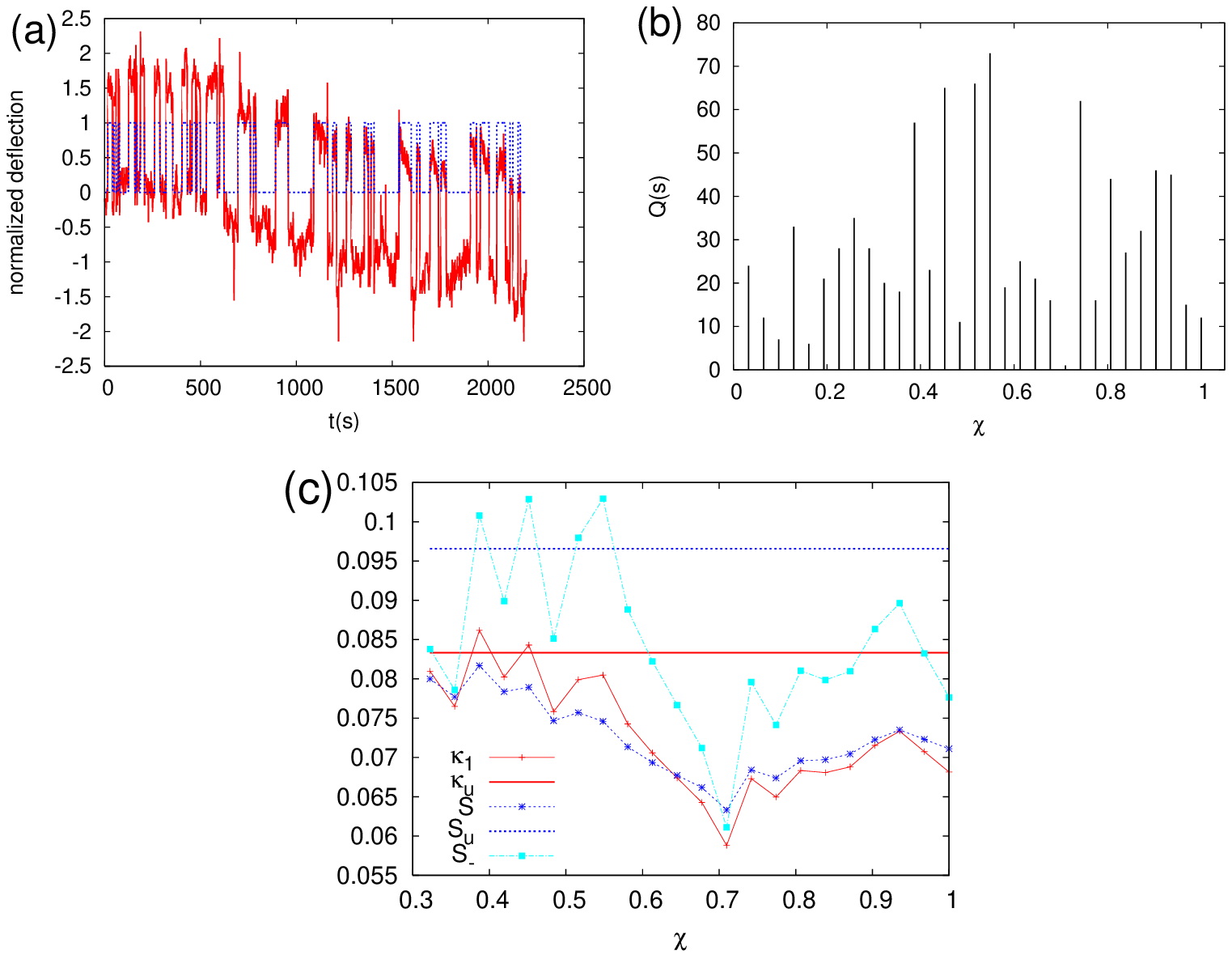}
\caption{(color) The electric signal recorded at KER on
February 11, 2007 (red) along with its dichotomous representation
which is marked by the dotted (blue) line. (b) How the signal in
(a) is read in natural time. (c) The values of $\kappa_1$, $S$ and
$S_-$ as the signals evolves from $\chi=0$ to $\chi=1$ versus the
natural time $\chi$. Note that upon the completion of the signal,
i.e., at $\chi=1$, the $\kappa_1$ value is close to 0.07, whereas
both $S$ and $S_-$ are smaller than $S_u$. For the reader's
convenience, the red and blue horizontal lines show the values
$\kappa_u=1/12$ and $S_u=0.0966$ of $\kappa_1$ and $S$,
respectively that correspond to a ``uniform'' ($u$) distribution.}
\label{f4}
\end{figure*}

We now proceed to the presentation of the most recent experimental
data of SES activities recorded in Greece by means of the
procedure described in detail in Refs.\cite{VAR91,VAR93,newbook}.
Figure \ref{f3} depicts four electrical disturbances that have
been recently recorded at three measuring stations termed Lamia
(LAM), located at $\approx 150$km north-west of Athens, Patras
(PAT) $\approx 160$km west of Athens,  and Keratea (KER) $\approx
30$km east south-east of Athens.  The signals are presented here
in normalized units, i.e., subtracting the mean value and dividing
the result by the standard deviation. The two electric signals in
Fig.\ref{f3}(a)  were recorded at LAM on November 13, 2006 and
January 8, 2007, respectively, the latter being of larger actual
amplitude than the former. In the upper panel of Fig.\ref{f3}(b)
an electric signal recorded at PAT on February 8, 2007 is
depicted, whereas  in the lower panel  we also insert -for the
sake of comparison- the signal that was recorded at the same
station almost one year ago ,i.e., on February 13, 2006. This
recent signal, i.e., the one on February 8, 2007,  has an
amplitude $\approx 70$\% larger\cite{Zaky} than that recorded one
year ago which has been analyzed in Ref.\cite{NAT06B}. Its subsequent seismic activity in discussed in the Appendix A (while more recent SES activities are presented in Appendix B). Finally, in
Fig.\ref{f3}(c) we depict a signal recorded at  KER on February
11, 2007.

All these four recent signals were analyzed in natural time and
found to be consistent with the conditions (\ref{eq1}) and
(\ref{eq2}), thus they can be classified\cite{NAT06A,NAT06B} as
SES activities. For example, in the lowest panel of Fig.\ref{f4}
we depict the evolution of the parameters $\kappa_1$, $S$ and
$S_-$ for one of these signals, i.e., the one recorded at KER on
February 11, 2007. The leftmost panel of this figure shows the
original time series (along with its dichotomous presentation),
while the rightmost panel indicates how the signal is read in
natural time. Note,  in Fig.\ref{f4}(c), that at $\chi=1$ (i.e.,
upon the completion of the signal) $\kappa_1$ reaches the value
$\kappa_1=0.068 \pm 0.003$, while $S$ and $S_-$ are $S=0.071 \pm
0.003$ and $S_-=0.078 \pm 0.003$. Thus, we have assured that,
within the experimental uncertainty, $\kappa_1 \approx 0.07$ and
$S,S_-< S_u$. Furthermore, upon shuffling the $Q_k$ randomly, we
have found that, in all these four signals, the variance
$\kappa_1$ and the entropies $S, S_-$ turn to be equal to the
values expected from a ``uniform'' distribution, which
assures\cite{NAT06B} that their self-similarity solely stems from
{\em temporal} correlations.

\section{Conclusions}
In summary, using the newly introduced concept of natural time:(a)
a simple model is proposed that exhibits $1/f^a$ behavior with $a$
close to unity and (b) electric signals, recorded during the last
few months in Greece, are classified as SES activities since they
exhibit {\em infinitely} ranged temporal correlations. This
excludes any possibility of attributing these signals to nearby
man-made sources, because the latter have {\em weaker} temporal
correlations (i.e., their Hurst exponent lies usually in the range
0.5 to 0.75\cite{NAT03,NAT03B}).

\appendix

\section{What happened after the SES activity at PAT  on February 8, 2007, until April 23, 2007.}

   According to the Athens observatory (the seismic data of which
will be used here),  a series of strong earthquakes (EQs) with
magnitudes ranging from 5.0 to 6.0-units  occurred as follows:
First, a 6.0 EQ at Kefallonia area, i.e., $38.34^o$N $20.42^o$E,
at 13:57 UT on March 25, 2007. Second, a cluster of four magnitude
class 5.0  EQs on April 10, 2007 with an epicenter close to
Trichonida lake, i.e., around $38.5^o$N $21.6^o$E.

We show that the occurrence time of the impending strong EQ
activity can be estimated by following the procedure described in
Refs.\cite{NAT01,newbook,VAR05C,NAT06B,EPAPS}, as it was indicated in
Ref.\cite{Zaky}. (We clarify that, during the last decade,
preseismic information based on SES activities is issued {\em
only} when the magnitude of the strongest EQ of the impending EQ
activity is estimated to be -by means of the SES
amplitude[38-40,52,53]- comparable to 6.0 units or
larger\cite{newbook}.)

\begin{figure*}
\includegraphics{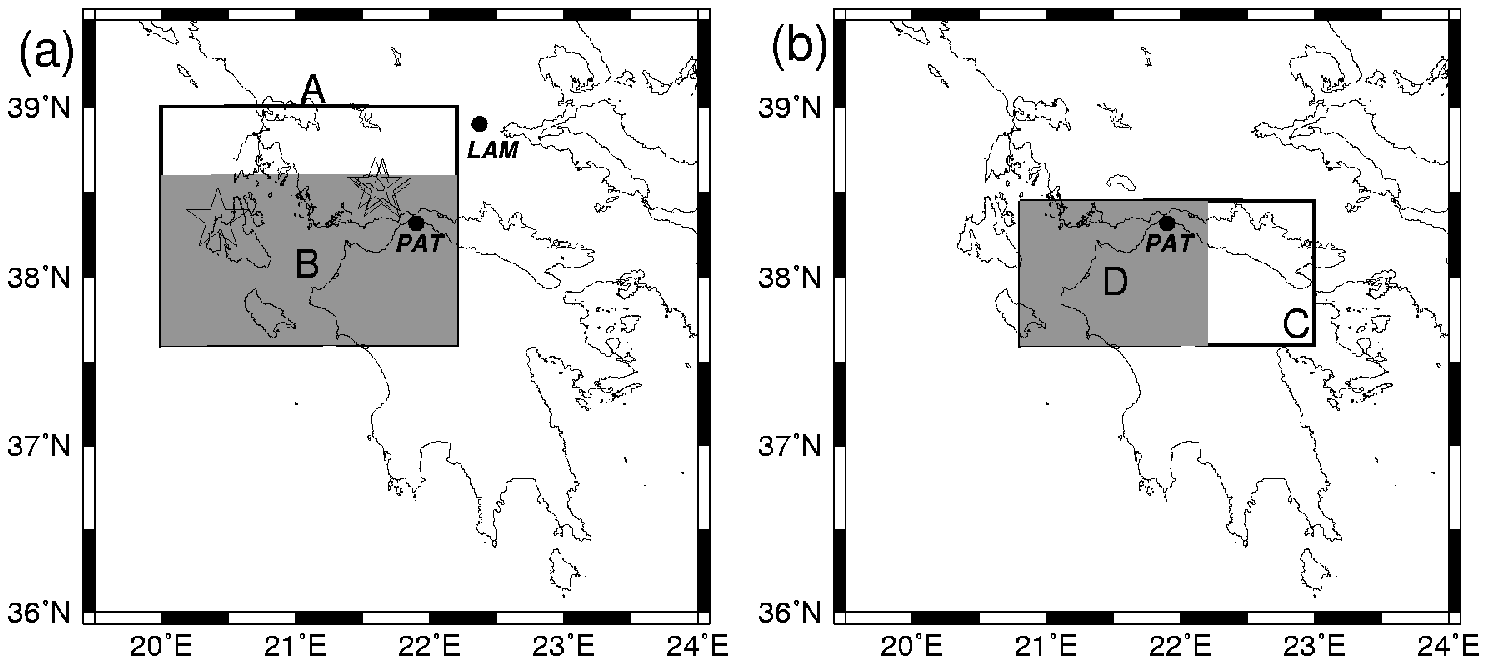}
\caption{The maps in (a) and (b) show the areas A,B and C,D, respectively. The stars in A stand for the strong 6.0 EQ that occurred on March 25, 2007 in Kefallonia (leftmost star) and the four magnitude 5.0-class EQs on April 10, 2007 close to Trichonida lake.}
\label{af1}
\end{figure*}

\begin{figure*}
\includegraphics{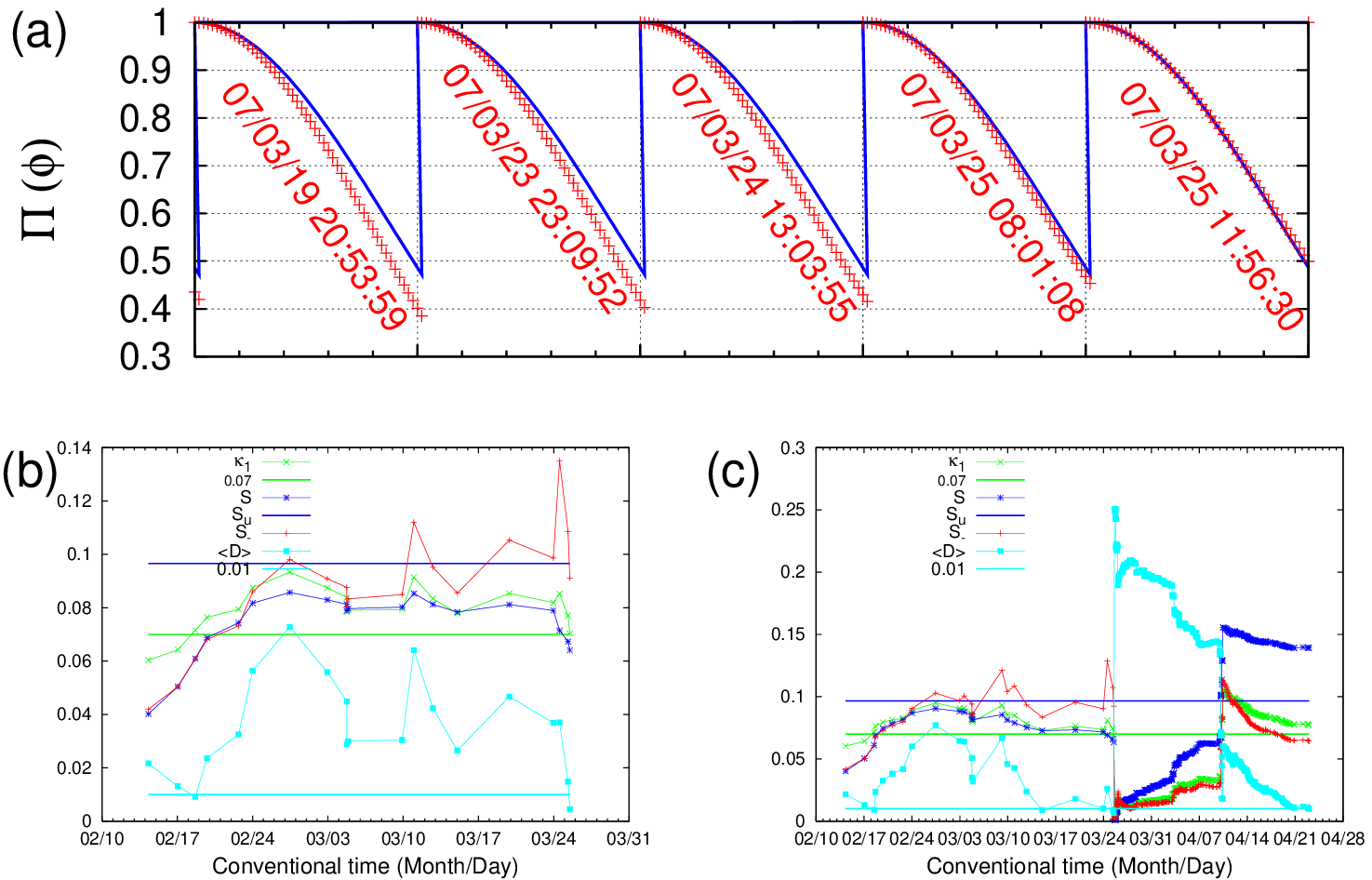}
\caption{(color) (a) The normalized power spectrum(red) $\Pi (\phi
)$ of the seismicity   as it evolves event by event (whose date
and time (UT) of occurrence are written in each panel) after the
initiation of the SES activity on February 8, 2007.  The
excerpt presented here refers to the period 19 to 25 March, 2007
and corresponds to the area B $M_{thres}=3.2$.  In each case only
the spectrum in the area $\phi \in [0,0.5]$ is depicted (separated
by the vertical dotted lines), whereas the  $\Pi (\phi )$ of
Eq.(2) is depicted by blue color. The minor horizontal ticks for
$\phi$ are marked every 0.1. (b), (c) Evolution of the
parameters  $\langle D \rangle$, $\kappa_1$, $S$ and $S_{-}$ after the initiation of the SES activity on February 8, 2007 for the areas B ($M_{thres}=3.2$) and 
A($M_{thres}=3.2$), respectively. In (b) the period just before the 6.0 EQ is shown, whereas in  (c) the period is 
extended until 03:26:33.4 UT on April 23, 2007.}
\label{af2}
\end{figure*}

We study how the seismicity evolved after the recording of the SES
activity at PAT on February 8, 2007, by considering either the area
A:$N_{37.6}^{39.0}E_{20.0}^{22.2}$ or its smaller area
B:$N_{37.6}^{38.6}E_{20.0}^{22.2}$, which surround the EQ
epicenters and the PAT station (see Fig.\ref{af1}(a)). If we set the
natural time for seismicity zero at the initiation of the  SES
activity on February 8, 2007, we form time series of seismic
events in natural time for various time windows as the number $N$
of consecutive (small) EQs increases. We then compute the
normalized power spectrum in natural time $\Pi (\phi )$ for each
of the time windows. Excerpt of these results, which refers to the
values deduced during the period from 20:53:59 UT on March 19 to 11:56:30 UT on 25 March, 2007, is
depicted in red in Fig.\ref{af2}(a).  This figure
corresponds to the area B  with magnitude threshold (hereafter
referring to the local magnitude ML or the `duration' magnitude
MD) $M_{thres}=3.2$. In the same figure, we plot in blue the power
spectrum obeying the relation (2) which holds, as mentioned, when
the system enters the {\em critical} stage ($\omega = 2\pi \phi$,
where $\phi$ stands for the natural frequency). The date and the
time of the occurrence of each small earthquake (with magnitude
exceeding (or equal to) the aforementioned threshold) that
occurred in area  B, is also written in red in
each panel. An inspection of this figure reveals that the red line
approaches the blue line as $N$ increases and a {\em coincidence}
occurs at the last small event which had a magnitude 3.2 and
occurred at 11:56:30 UT on March 25, 2007, i.e., just two hours
before the strong 6.0 EQ. To ensure that this coincidence is a
{\em true} one (see also below) we also calculate the evolution of
the quantities $\kappa_1$,$S$ and $S_{-}$   and the results are
depicted in Fig. \ref{af2}(b) and \ref{af2}(c) for the same magnitude thresholds  for each of the areas B and A, respectively.

The conditions for a coincidence to be considered as {\em true}
are the following (e.g., see Refs. [30, 48, 49, 55,56]): First, the
`average' distance $\langle D \rangle$ between the empirical and
the theoretical $\Pi(\phi )$(i.e., the red and the blue line,
respectively, in Fig.\ref{af2}(a)) should be smaller than $10^{-2}$. See
Fig. \ref{af2}(b),(c) where we plot $\langle D \rangle$ versus the
conventional time for the aforementioned two areas B and A, respectively. Second, in the examples observed to date, a few events
{\em before} the coincidence leading to the strong EQ, the
evolving $\Pi(\phi )$ has been found to approach that of the
relation (2), i.e., the blue one in Fig.\ref{af2}(a) , from {\em below}
(cf. this reflects that during this approach the $\kappa_1$-value
decreases as the number of events increases). In addition, both
values $S$ and $S_{-}$ should be smaller than $S_u$ at the
coincidence. Finally, since the process concerned is self-similar
({\em critical} dynamics), the time of the occurrence of the
(true) coincidence should {\em not} change, in principle, upon
changing the (surrounding) area
used in the calculation. Note that in Fig. \ref{af2}(b), upon the
occurrence of the aforementioned last small event at 11:56:30 UT of March 25,
2007, in area B  the $\langle D \rangle$ value becomes
smaller than $10^{-2}$. The same was found to hold for the area A (see Fig.\ref{af2}(c)).

An inspection of Fig. \ref{af2}(c) shows that upon the occurrence
of the four magnitude class 5.0 EQs on April 10, 2007, the parameters $\kappa_1$,
$S$, and $S_-$ jumped to values larger than those of the
``uniform'' distribution and then started to decrease again. Among
these parameters, however, $S$ still remains appreciably higher
than $S_u$, while $S_-$ it already became smaller than $S_u$ and
$\kappa_1$ gradually approaches 0.070.

\begin{figure*}
\includegraphics{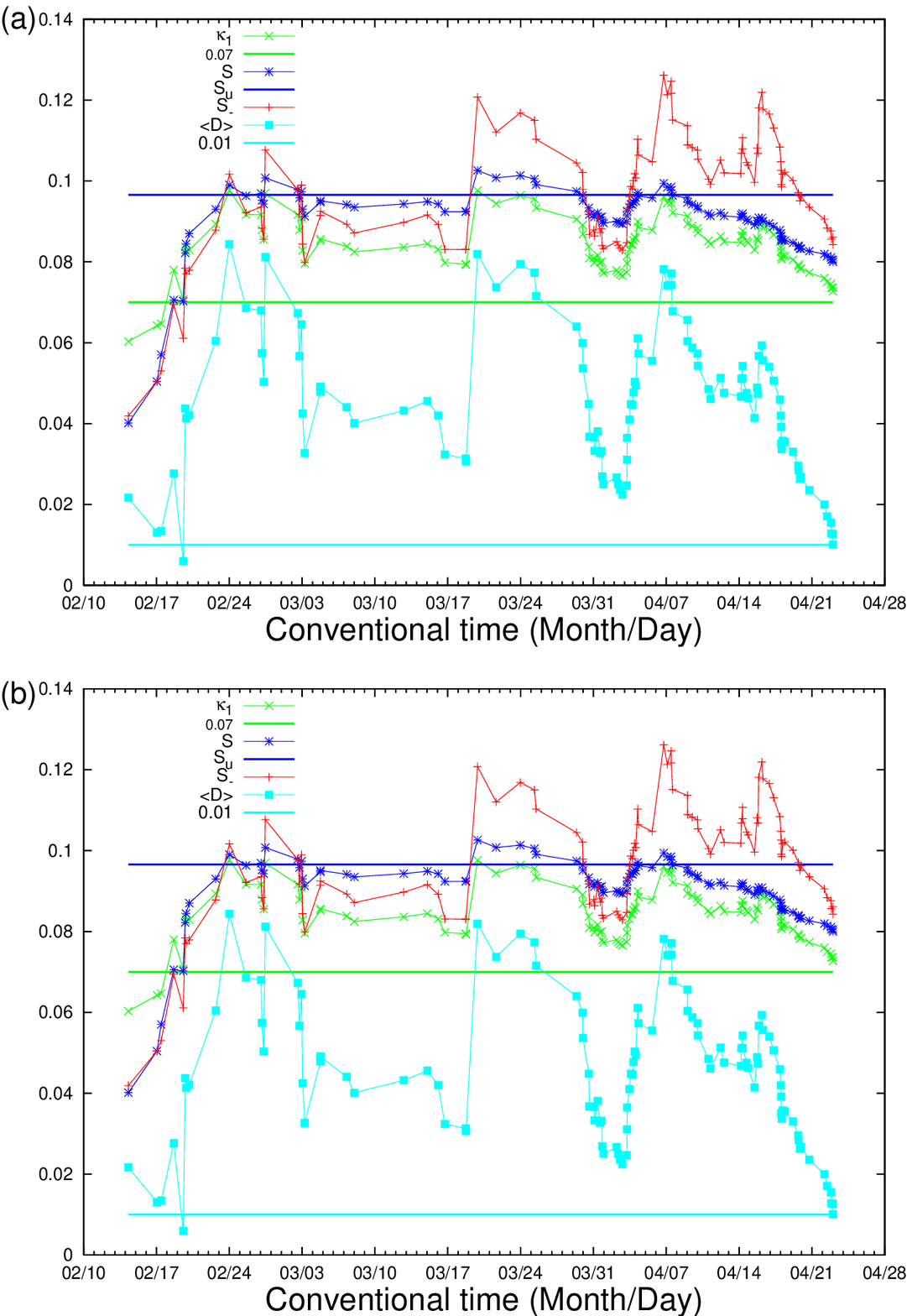}
\caption{(color) (a) and (b) depict the evolution of the
parameters  $\langle D \rangle$, $\kappa_1$, $S$ and $S_{-}$ after the initiation of the SES activity on February 8, 2007, but for the areas C  and 
D, respectively, until 00:49:12 UT on April 23, 2007.}
\label{af3}
\end{figure*}

In order to elucidate the situation, we  proceeded to the
additional study of two areas C and D, which did {\em not} include
the epicenters of the EQs at Kefallonia and Trichonida lake, i.e.,
C: $N_{37.6}^{38.45}E_{20.8}^{23.0}$ and its smaller region D:
$N_{37.6}^{38.45}E_{20.8}^{22.2}$. The corresponding parameters
are depicted in Fig.\ref{af3}, which exhibit the following interesting 
feature: They gradually decrease, thus approaching the aforementioned values that
characterize the proximity to the critical point (CP). The occurrence of the next few small events in these areas
will reveal whether CP will be finally reached or not. This study is
still in progress.

\begin{figure*}
\includegraphics{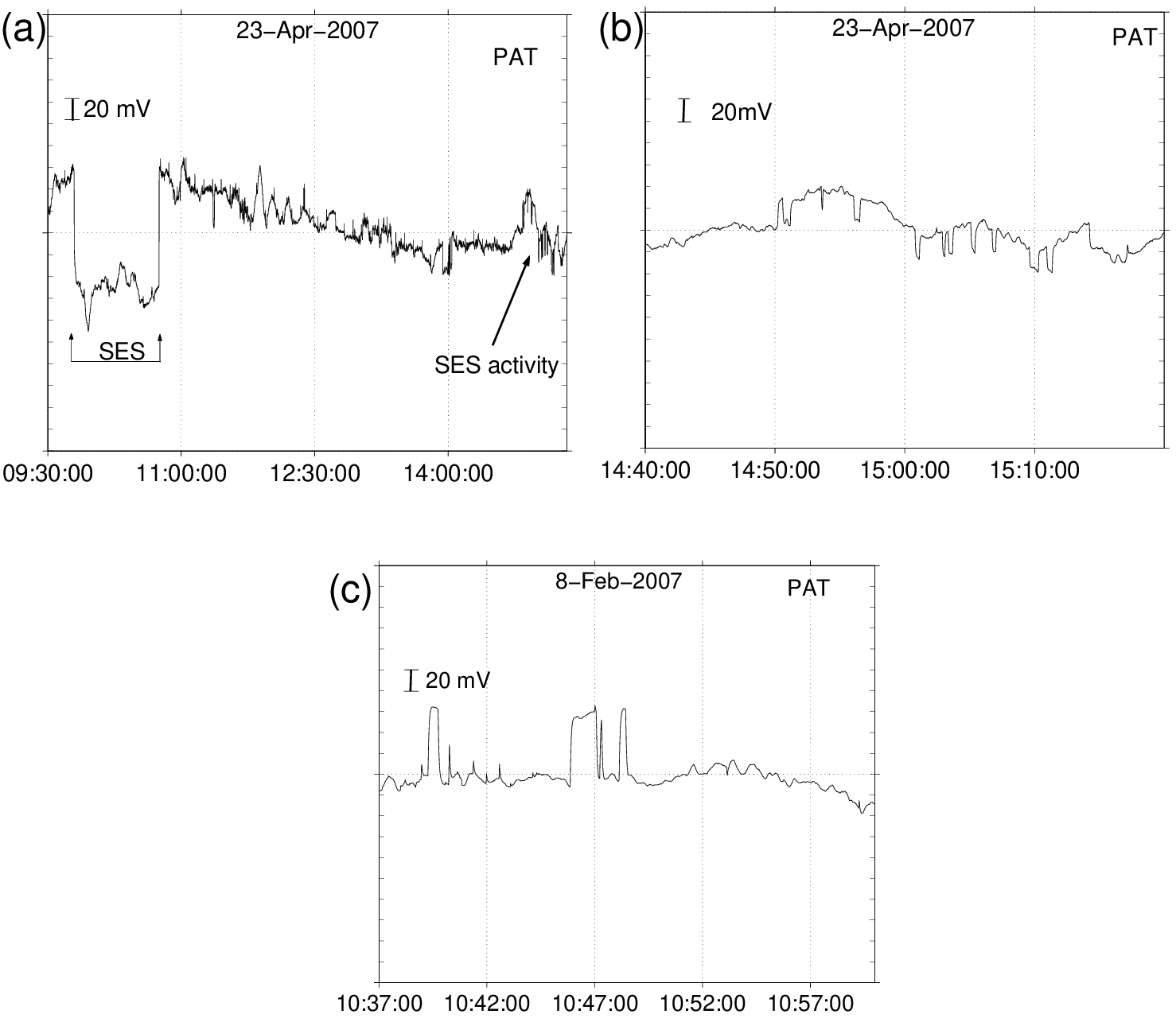}
\caption{(a): Electrical recordings at PAT on April 23, 2007. (b): Excerpt of (a), in an expanded time scale, to show the SES activity indicated by the arrow in (a). (c): For the sake of comparison, the SES activity recorded by the same station on February 8, 2007, is depicted.}
\label{af4}
\end{figure*}

The interesting feature seen in Fig. \ref{af3} seems to conform with the
following fact: Just on April 23, 2007, a {\em single} SES (in the
sense discussed in Ref. 52) was recorded (Fig. \ref{af4}(a)) which was
followed by an SES activity (Fig. \ref{af4}(b)). Comparing the actual
amplitude of this single SES to that of the SES activity recorded on February 8, 2007
(Fig. \ref{af4}(c)), we find that it is evidently stronger.

\section{Further experimental data added on May 8,2007.}

\begin{figure*}
\includegraphics{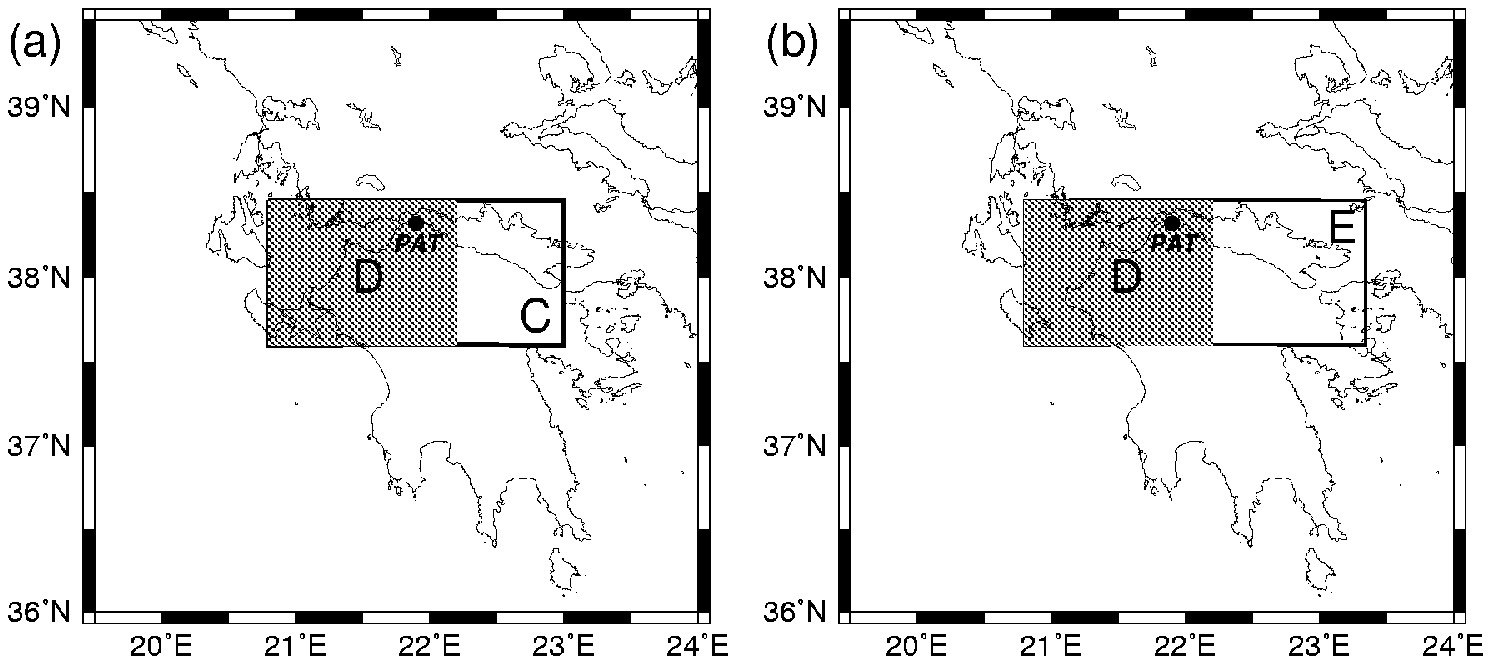}
\caption{The maps in (a) and (b) show the areas C,D and D,E, respectively. The map in (a) is a reproduction of Fig.\ref{af1}(b) (originally submitted on April 24, 2007) which also includes the 5.1 EQ (star) on May 6, 2007, close to Zakynthos Island. In (b) the area F: N$^{38.45}_{37.6}$E$^{23.3}_{21.75}$ (not shown) is also currently studied.}
\label{af5}
\end{figure*}

At 01:34 UT on May 7,2007 an EQ of magnitude 5.1 occurred with an epicenter at $37.62^o$N$21.08^o$E depicted with a star in Fig.\ref{af5}(a). It actually lies within the areas D,C (Fig.\ref{af1}(b)) which exhibited the interesting feature discussed in advance in Fig.\ref{af3}. 

 \begin{figure*}
\includegraphics{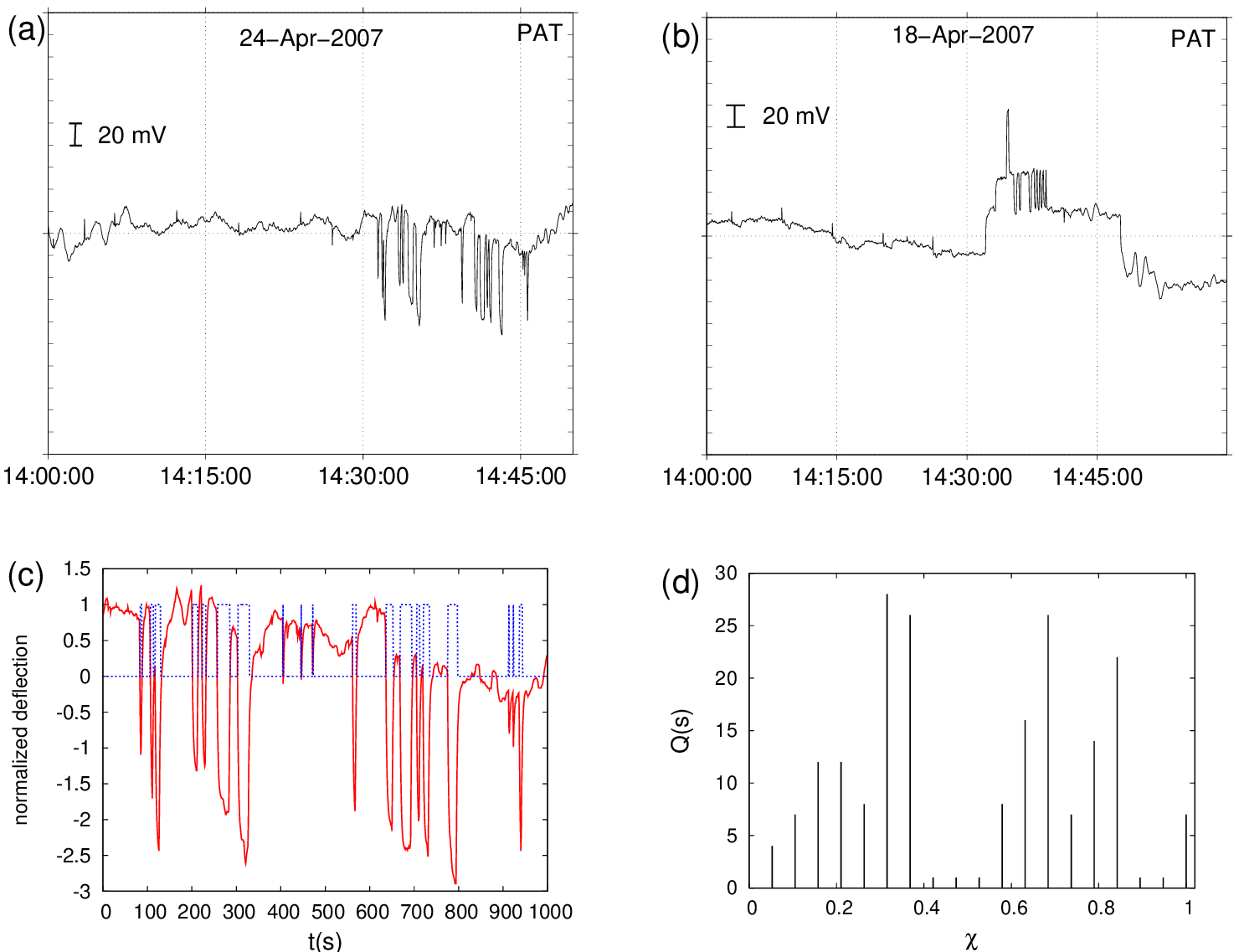}
\caption{(color) (a),(b): Electrical recordings at PAT on April 24, 2007, and April 18, 2007 respectively. (c): The electric signal depicted in (a) in normalized units  along with its dichotomous representation
which is marked by the dotted (blue) line. (d): How the signal in
(c) is read in natural time.}
\label{af6}
\end{figure*}

\begin{figure*}
\includegraphics{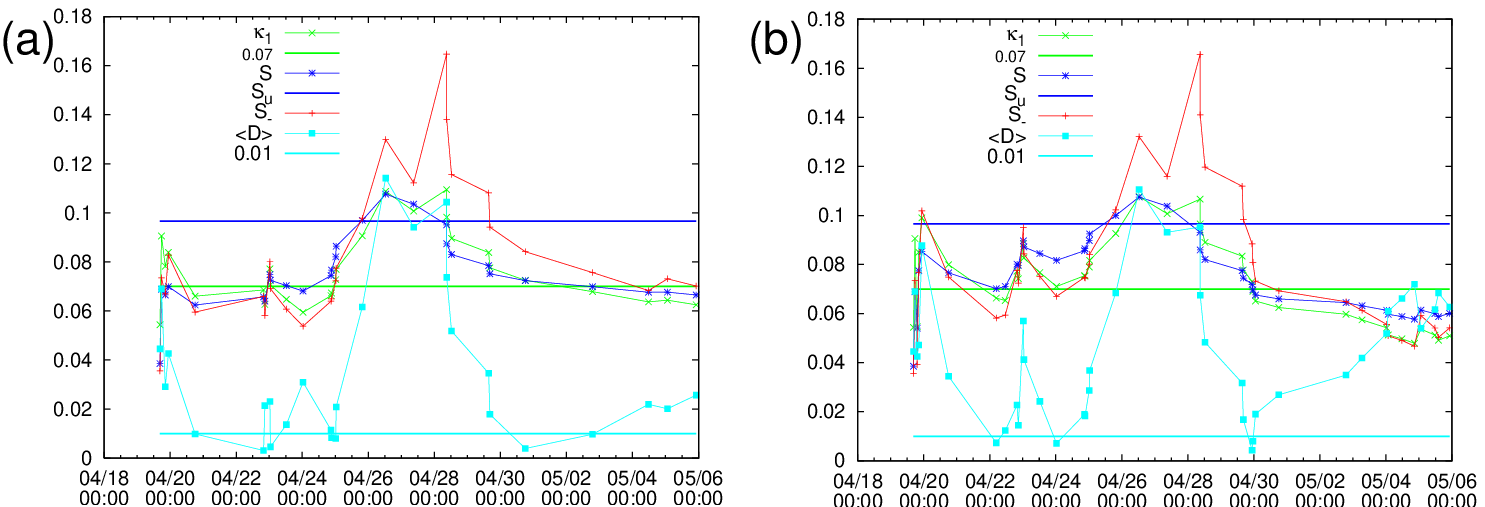}
\caption{(color) (a) and (b) depict the evolution of the
parameters  $\langle D \rangle$, $\kappa_1$, $S$ and $S_{-}$ after the initiation of the SES activity on April 18, 2007, but for the areas D  and 
E, respectively, until 04:14:12 UT on May 6, 2007.}
\label{af7}
\end{figure*}

Concerning the electrical data: Beyond the electrical recordings on April 23, 2007 depicted in Fig.\ref{af4}(a),(b), two additional electric signals have been recorded at PAT on April 24, 2007 (Fig.\ref{af6}(a)) and April 18, 2007 (Fig.\ref{af6}(b)); which have an amplitude markedly larger than that on February 13, 2006 (see also Ref.\cite{Zaky}). These have been classified as SES activities after analyzing them in natural time and applying the relevant criteria. For example, if we read in natural time the signal on April 24, 2007 (Fig.\ref{af6}(d)) -the dichotomous representation of which is marked by the dotted (blue) line in Fig.\ref{af6}(c)- we find the values $\kappa_1= 0.067\pm 0.003$, $S= 0.072 \pm 0.003$, $S_-=0.069\pm 0.003$ which do obey the conditions (3) and (4). 

We now investigate the seismicity after the aforementioned three SES activities on April 18, April 23 and April 24, 2007. The investigation is made in the areas D and E (Fig.\ref{af5}(b)), the latter being almost equal to the former region C but slightly extended to the east, i.e., E: N$^{38.45}_{37.6}$E$^{23.3}_{20.8}$. Starting the computation of seismicity from the initiation of the SES activity on April 18, 2007, we obtain the results depicted in Figs.\ref{af7}(a) and \ref{af7}(b) for the areas D and E, respectively. An inspection of the parameters $\langle D \rangle$, $\kappa_1$,$S$ and $S_-$ reveals that they exhibited a {\em true} coincidence (as discussed in Appendix A) around April 30, 2007, i.e., around one week before the 5.1 EQ on May 7, 2007 mentioned above. This study of the evolution of seismicity in the areas D, E still continues.

In view of the fact that three SES activities (April 18, 23 and 24, 2007) have been observed with different polarities (see Fig.\ref{af4}(a) and Figs.\ref{af6}(a),(b)) while only one strong EQ has occurred to date, our  investigation is also currently extended to the area F: N$^{38.45}_{37.6}$E$^{23.3}_{21.75}$ (which is a sub-region of E to the east of PAT). This study is still in progress in order to discriminate whether this area may approach CP or not. 

\section{Further data until May 14,2007.}

\begin{figure*}
\includegraphics{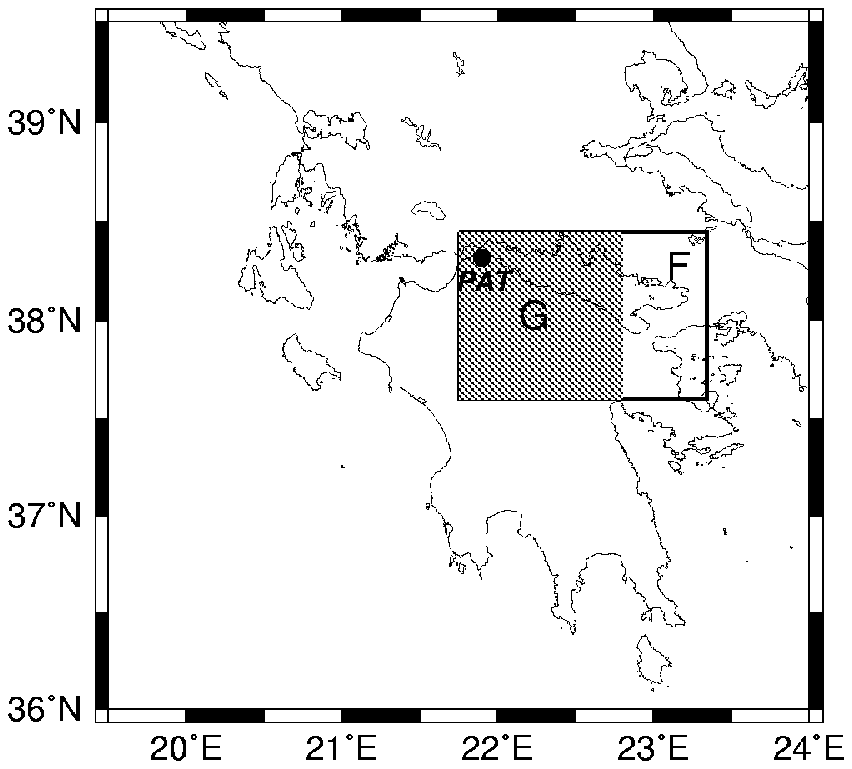}
\caption{A map showing  the areas G and F.}
\label{cf1}
\end{figure*}

\begin{figure*}
\includegraphics{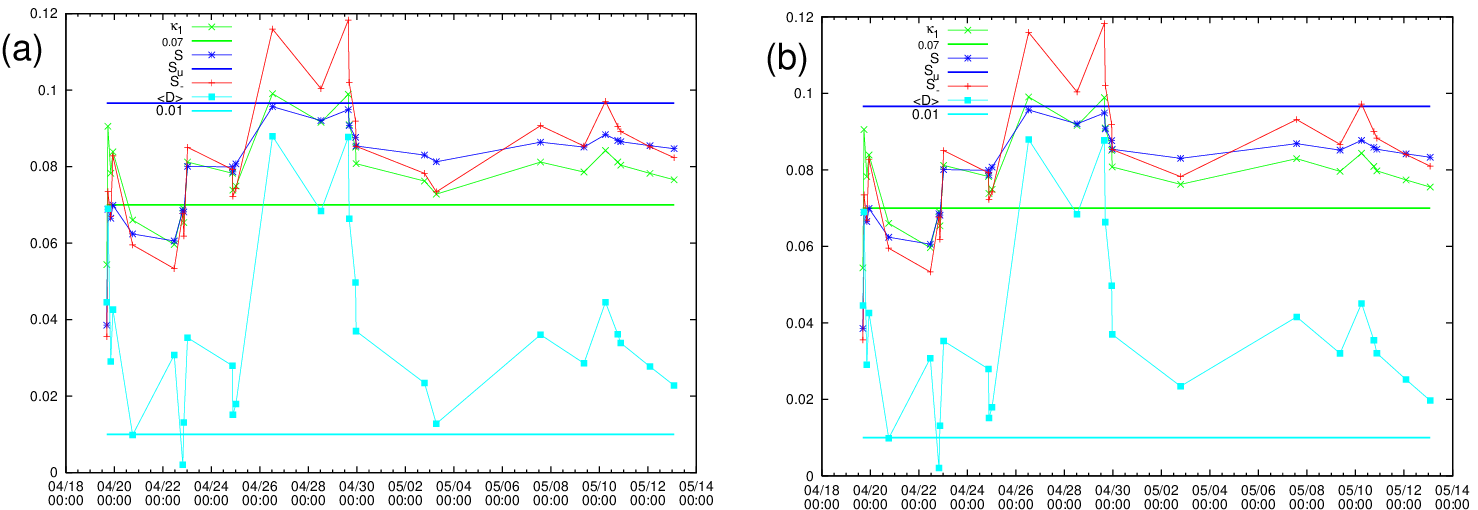}
\caption{(color) (a) and (b) depict the evolution of the
parameters  $\langle D \rangle$, $\kappa_1$, $S$ and $S_{-}$ after the initiation of the SES activity on April 18, 2007, but for the areas F  and 
G, respectively, for $M_{thres} \geq 3.0$ until 02:22:17 UT  on May 14, 2007.}
\label{cf2}
\end{figure*}

As mentioned in the previous Appendix, we {\em also} currently
study the area F:N$_{37.6}^{38.45}$ E$_{21.75}^{23.34}$ shown in
Fig.\ref{cf1}. Starting the calculation from the initiation of the SES on April 18, 2007
(Fig.\ref{af6}(b)) the resulting parameters (until 02:22:17 UT on May 14, 2007) $\langle D \rangle$, $\kappa_1$, $S$, and $S_-$ from the computation of
seismicity are shown in Fig.\ref{cf2}(a), for $M_{thres} \geq 3.0$. To
investigate the spatial invariance, we also give in Fig.\ref{cf2}(b) the
relevant results for a subregion G of F, i.e., G:N$_{37.6}^{38.45}$
E$_{21.75}^{22.8}$. Interestingly, both Figs. \ref{cf2}(a), (b) exhibit a
systematic tendency to gradually approach CP. The extent to which
they will finally obey the aforementioned conditions for a {\em true} coincidence or not,
can be judged upon the occurrence of the next few events in the areas F and G under discussion.


\begin{thebibliography}{10}
\expandafter\ifx\csname url\endcsname\relax
  \def\url#1{\texttt{#1}}\fi
\expandafter\ifx\csname
urlprefix\endcsname\relax\def\urlprefix{URL }\fi

\bibitem{MAN99}
B.~B. Mandelbrot, Multifractals and 1/f Noise, Springer-Verlag,
New York, 1999.

\bibitem{ANT02}
T.~Antal, M.~Droz, G.~Gy{\"o}rgyi, Z.~R\'{a}cz, Roughness
distributions for
  1/f$^a$ signals, Phys. Rev. E 65 (2002) 046140.

\bibitem{WEI88}
M.~B. Weissman, 1/f noise and other slow, nonexponential kinetics
in condensed
  matter, Rev. Mod. Phys. 60 (1988) 537--571.

\bibitem{MUS76}
T.~Musha, H.~Higuchi, 1/f fluctuation of a traffic current on an
expressway,
  Jpn. J. Appl. Phys. 15 (1976) 1271--1275.

\bibitem{NAG95}
K.~Nagel, M.~Paczuski, Emergent traffic jams, Phys. Rev. E 51
(1995)
  2909--2918.

\bibitem{ZHA95}
X.~Zhang, G.~Hu, 1/f noise in a two-lane highway traffic model,
Phys. Rev. E 52
  (1995) 4664--4668.

\bibitem{NAK97}
A.~Nakahara, T.~Isoda, 1/f$^a$ density fluctuations at the
slugging
  transition point of granular flows through a pipe, Phys. Rev. E 55 (1997)
  4264--4273.

\bibitem{gol02}
A.~L. Goldberger, L.~A.~N. Amaral, J.~M. Hausdorff, P.~C. Ivanov,
C.-K. Peng,
  H.~E. Stanley, Fractal dynamics in physiology: Alterations with disease and
  aging, Proc. Natl. Acad. Sci. USA 99 (2002) 2466--2472.

\bibitem{PEN93}
C.-K. Peng, J.~Mietus, J.~M. Hausdorff, S.~Havlin, H.~E. Stanley,
A.~L.
  Goldberger, Long-range anticorrelations and non-Gaussian behavior of the
  heartbeat, Phys. Rev. Lett. 70 (1993) 1343--1346.

\bibitem{MER99}
S.~Mercik, K.~Weron, Z.~Ziwy, Statistical analysis of ionic
current
  fluctuations in membrane channels, Phys. Rev. E 60 (1999) 7343--7348.

\bibitem{MAN69b}
B.~Mandelbrot, J.~R. Wallis, Some long run properties of
geophysical records,
  Water Resourc. Res. 5 (1969) 321--340.

\bibitem{LIL00}
F.~Lillo, R.~N. Mantegna, Variety and volatility in financial
markets, Phys.
  Rev. E 62 (2000) 6126--6134.

\bibitem{GOM05}
J.~M.~G. C\'{o}mez, A.~Rela\~{n}o, J.~Retamosa, E.~Faleiro,
L.~Salasnich,
  M.~Vrani\v{c}ar, M.~Robnik, 1/f$^a$ noise in spectral fluctuations of
  quantum systems, Phys. Rev. Lett 94 (2005) 084101.

\bibitem{REL02}
A.~Rela\~{n}o, J.~M.~G. C\'{o}mez, R.~A. Molina, J.~Retamosa,
E.~Faleiro,
  Quantum chaos and 1/f noise, Phys. Rev. Lett 89 (2002) 244102.

\bibitem{SAN05}
M.~S. Santhanam, J.~N. Bandyopadhyay, Spectral fluctuations and
1/f noise in
  the order-chaos transition regime, Phys. Rev. Lett. 95 (2005) 114101.

\bibitem{SAN06}
M.~S. Santhanam, J.~N. Bandyopadhyay, D.~Angom, Quantum spectrum
as a time
  series: Fluctuation measures, Phys. Rev. E 73 (2006) 015201.

\bibitem{PRE78}
W.~H. Press, Flicker noises in astronomy and elsewhere, Comments
Astrophys. 7
  (1978) 103.

\bibitem{GIL95}
D.~L. Gilder, T.~Thornton, M.~W. Mallon, 1/f noise in human
cognition, Science
  267 (1995) 1837--1839.

\bibitem{YOS00}
H.~Yoshinaga, S.~Miyazima, S.~Mitake, Fluctuation of biological
rhythm in
  finger tapping, Physica A 280 (2000) 582--586.

\bibitem{BER63}
J.~M. Berger, B.~B. Mandelbrot, A new model for the clustering of
errors on
  telephone circuits, IBM J. Res. Dev. 7 (1963) 224--236.

\bibitem{KOG96}
S.~Kogan,{\em Electronic Noise and Fluctuations in Solids},
{Cambridge University
  Press}, {Cambridge}, 1996.

\bibitem{COL00}
P.~G. Collins, M.~S. Fuhrer, A.~Zettl, 1/f noise in carbon
nanotubes, Appl.
  Phys. Lett. 76 (2000) 894--896.

\bibitem{KIS97}
L.~B. Kiss, {\em et al.}, Diffusive fluctuations, long-time and
short-time
  cross-correlations in the motion of vortice-pancakes in different layers of
  YBCO/PBCO superlattices, Solid State Commun. 101 (1997) 51--56.

\bibitem{SOR00}
D.~Sornette, {\em Critical Phenomena in the Natural Sciences:
Chaos, Fractals,
  Selforganization, and Disorder: Concepts and Tools}, Springer-Verlag, Berlin,
  2000.

\bibitem{YAK00}
A.~V. Yakimov, F.~N. Hooge, A simple test of the gaussian
character of noise,
  Physica B-condensed matter 291 (2000) 97--104.

\bibitem{BAK87}
P.~Bak, C.~Tang, K.~Wiesenfeld, Self-organized criticality: An
explanation of
  the 1/f noise, Phys. Rev. Lett. 59 (1987) 381.

\bibitem{BAK96}
P.~Bak, {\em How Nature Works}, {Copernicus}, {New York}, 1996.

\bibitem{ANT01}
T.~Antal, M.~Droz, G.~Gy{\"o}rgyi, Z.~R\'{a}cz, 1/f noise and
extreme value
  statistics, Phys. Rev. Lett 87 (2001) 240601.

\bibitem{DAV02}
J.~Davidsen, H.~G. Schuster, Simple model for 1/f$^a$ noise, Phys.
Rev. E 65
  (2002) 026120.

\bibitem{NAT01}
P.~A. Varotsos, N.~V. Sarlis, E.~S. Skordas, Spatio-temporal
comlpexity aspects
  on the interrelation between seismic electric signals and seismicity,
  Practica of Athens Academy 76 (2001) 294--321.

\bibitem{NAT02}
P.~A. Varotsos, N.~V. Sarlis, E.~S. Skordas, Long-range
correlations in the
  electric signals the precede rupture, Phys. Rev. E 66 (2002) 011902.

\bibitem{NAT02A}
P.~Varotsos, N.~Sarlis, E.~Skordas, Seismic electric signals and
seismicity: On
  a tentative interrelation between their spectral content, Acta Geophys. Pol.
  50 (2002) 337--354.

\bibitem{NAT03}
P.~A. Varotsos, N.~V. Sarlis, E.~S. Skordas, Long-range
correlations in the
  electric signals the precede rupture: Further investigations, Phys. Rev. E 67
  (2003) 021109.

\bibitem{NAT03B}
P.~A. Varotsos, N.~V. Sarlis, E.~S. Skordas, Attempt to
distinguish electric
  signals of a dichotomous nature, Phys. Rev. E 68 (2003) 031106.

\bibitem{NAT04}
P.~A. Varotsos, N.~V. Sarlis, E.~S. Skordas, M.~S. Lazaridou,
Entropy in
  natural time domain, Phys. Rev. E 70 (2004) 011106.

\bibitem{NAT05}
P.~A. Varotsos, N.~V. Sarlis, E.~S. Skordas, M.~S. Lazaridou,
Natural entropy
  fluctuations discriminate similar-looking electric signals emitted from
  systems of different dynamics, Phys. Rev. E 71 (2005) 011110.

\bibitem{ABE05}
S.~Abe, N.~V. Sarlis, E.~S. Skordas, H.~K. Tanaka, P.~A. Varotsos,
Origin of
  the usefulness of the natural-time representation of complex time series,
  Phys. Rev. Lett. 94 (2005) 170601.

\bibitem{proto}
P.~Varotsos, K.~Alexopoulos, Physical properties of the variations
of the
  electric field of the earth preceding earthquakes, I and II, Tectonophysics 110
  (1984) 73--98; {\em ibid.} 99--125.

\bibitem{var86b}
P.~Varotsos, K.~Alexopoulos, K.~Nomicos, M.~Lazaridou, Earthquake
prediction
  and electric signals, Nature (London) 322 (1986) 120.

\bibitem{var88x}
P.~Varotsos, K.~Alexopoulos, K.~Nomicos, M.~Lazaridou, Official
earthquake
  prediction procedure in Greece, Tectonophysics 152 (1988) 193--196.

\bibitem{var99}
P.~Varotsos, N.~Sarlis, M.~Lazaridou, Interconnection of defect
parameters and
  stress-induced electric signals in ionic crystals, Phys. Rev. B 59 (1999) 24.

\bibitem{grl}
N.~Sarlis, M.~Lazaridou, P.~Kapiris, P.~Varotsos, Numerical model
of the
  selectivity effect and $\Delta V/L$ criterion, Geophys. Res. Lett. 26 (1999)
  3245--3248.

\bibitem{varbook}
P.~Varotsos, K.~Alexopoulos, {\em Thermodynamics of Point Defects
and their Relation
  with Bulk Properties}, North Holland, Amsterdam, 1986.

\bibitem{VAR06PRB}
P.~Varotsos, N.~Sarlis, S.~Skordas, Flux avalanches in
$YBa_2Cu_3O_{7-x}$ films
  and rice piles: Natural time domain analysis, Phys. Rev. B 73 (2006) 054504.

\bibitem{AEG03}
C.~M. Aegerter, R.~Gunther, R.~J. Wijngaarden, Avalanche dynamics,
surface
  roughening, and self-organized criticality: Experiments on a
  three-dimensional pile of rice, Phys. Rev. E 67 (2003) 051306.

\bibitem{AEG04A}
C.~M. Aegerter, K.~A. Lorincz, M.~S. Welling, R.~J. Wijngaarden,
Extremal
  dynamics and the approach to the critical state: Experiments on a three
  dimensional pile of rice, Phys. Rev. Lett. 92 (2004) 058702.

\bibitem{NAT05B}
P.~A. Varotsos, N.~V. Sarlis, H.~K. Tanaka, E.~S. Skordas, Some
properties of
  the entropy in the natural time, Phys. Rev. E 71 (2005) 032102.

\bibitem{newbook}
P.~Varotsos, The Physics of Seismic Electric Signals, TERRAPUB,
Tokyo, 2005.

\bibitem{VAR05C}
P.~Varotsos, N.~Sarlis, H.~Tanaka, E.~Skordas, Similarity of
fluctuations in
  correlated systems: The case of seismicity, Phys. Rev. E 72 (2005) 041103.

\bibitem{NAT06A}
P.~A. Varotsos, N.~V. Sarlis, E.~S. Skordas, H.~K. Tanaka, M.~S.
Lazaridou,
   Entropy of seismic electric signals: Analysis in the natural time under time
  reversal, Phys. Rev. E 73 (2006) 031114.

\bibitem{ABE04}
U.~Tirnakli, S.~Abe, Aging in coherent noise models and natural
time, Phys.
  Rev. E 70 (2004) 056120.

\bibitem{VAR91}
P.~Varotsos, M.~Lazaridou, Latest aspects of earthquake prediction
in Greece
  based on seismic electric signals, Tectonophysics 188 (1991) 321--347.

\bibitem{VAR93}
P.~Varotsos, K.~Alexopoulos, M.~Lazaridou, Latest aspects of
earthquake
  prediction in Greece based on seismic electric signals,ii, Tectonophysics 224
  (1993) 1--37.


\bibitem{Zaky}
The SES activity on February 13, 2006, as discussed in
Ref.\cite{NAT06B}, was
  followed by a series of earthquakes, located at almost 100km WSW of PAT, the
  strongest of which was of magnitude class 6.0. The estimation of the time window of
  the impending earthquake can be made by computing the order parameter of
  seismicity following the procedure described in detail in
  Refs.\cite{NAT01,newbook,VAR05C}. 

\bibitem{NAT06B}
P.~A. Varotsos, N.~V. Sarlis, E.~S. Skordas, H.~K. Tanaka, M.~S.
Lazaridou,
  Attempt to distinguish long-range temporal correlations from the statistics
  of the increments by natural time analysis, Phys. Rev. E 74 (2006) 021123.
 
 \bibitem{EPAPS}
See EPAPS Document No. E-PLEEE8-73-134603 for additional
information.
  This document may be retrieved via the EPAPS homepage
  ({\tt http://www.aip.org/pubservs/epaps.html}) or from {\tt ftp.aip.org} in
  the directory /epaps/. See the EPAPS homepage for more information.
\end{thebibliography}

\end{document}